\documentclass[12pt]{article}
\usepackage{graphicx}
\usepackage{amsmath}
\usepackage{amssymb}
\usepackage[numbers,sort&compress]{natbib}
\title{The AB equations and the $\bar\partial$-dressing method in semi-characteristic coordinates
\footnotetext{}}
\author{Junyi Zhu\thanks{Email: jyzhu@zzu.edu.cn} and Xianguo Geng \\
{\small School of Mathematics and Statistics, Zhengzhou University,
Zhengzhou, Henan 450001, China}}
\date{}
\begin{document}
\maketitle
\begin{abstract}
The dressing method based on the $2\times2$ matrix $\bar\partial$-problem is generalized to study the canonical form of AB equations.
The soliton solutions for the AB equations are given by virtue of the properties of Cauchy matrix.
Asymptotic behaviors of the $N$-soliton solution are discussed.
\end{abstract}

\section{Introduction}
The AB equations have important applications in geophysical fluids and in nonlinear optics \cite{PJ70,PJ,MI,M-B,D-E-G-M}.
The important features are that the AB equations are integrable by the inverse scattering transform and can be reduced to the sine-Gordon equation\cite{G-J-M,G-M}.
The single-phase periodic solution is studied %
by the method for improving the effectiveness of one-phase periodic solutions of integrable equations
in \cite{K-P}, the envelope solitary wave and sine Waves are discussed in \cite{T-B}.
In addition, Guo {\em et al.} \cite{G-T} investigated the Painlev\'e property and conservation laws of one type of variable-coefficient AB equation,
and obtained the soliton solutions by Darboux transformation.

The $\bar\partial$-dressing method \cite{Z-M,B-M,BC,Z,SPM} is a powerful tools to construct and solve integrable nonlinear equations
as well as to describe their transformations and reductions. For a review see \cite{KBG,DL}, and references therein.

To our knowledge, The $N$-soliton solution of the AB equation has not been given and $\bar\partial$-dressing method for the AB equation is open.
In this paper, we study the AB equations in semi-characteristic coordinates by extended $\bar\partial$-dressing method \cite{Z-G} and
give their $N$-soliton solution.

The present paper is organized as follows. In Sec. \ref{sec:two}, the semi-characteristic coordinates $\xi$ and $\tau$
are introduced in the spectral transform matrix to derive the Lax pair of these equations, where the $\tau$-dependent linear spectral problem
is obtained by introducing a special singular dispersion relation. In Sec. \ref{sec:three}, suitable symmetry conditions
are applied to derive the AB equations in canonical form. In Sec. \ref{sec:four}, the properties of Cauchy matrix are used to discuss one-soliton, two-soliton,
as well as N-soliton solutions of the equations.
In the last section, we study the asymptotic behaviors of the $N$-soliton solution.

\setcounter{equation}{0}
\section{\label{sec:two}Spectral transform and Lax pair}
In this paper, we consider the $2\times2$ matrix $\bar{\partial}$-problem in the complex $k$-plane,
\begin{equation}\label{b1}
\bar{\partial}\psi(k,\bar{k})=\psi(k,\bar{k})R(k,\bar{k}),
\end{equation}
where $\bar\partial\equiv\partial/\partial\bar{k}$ and $R=R(k,\bar{k})$ is a spectral transform matrix which will be associated with a nonlinear equation.
It is readily verified that a solution of the $\bar{\partial}$-problem (\ref{b1}) with the canonical normalization
can be written as
\begin{equation}\label{b2}
\psi(k)=I+\psi RC_k,
\end{equation}
where $C_k$ denotes the Cauchy-Green integral operator acting on the left
$$\psi RC_k=\frac{1}{2i\pi}\iint\frac{{\rm d}z\wedge {\rm d}\bar{z}}{z-k}\psi(z)R(z),$$
and here we have suppressed the variable $\bar{k}$ dependence in $\psi$ and $R$.
It is readily verified that, for some matrix functions $f(k)$ and $g(k$), the operator $C_k$ satisfies
\begin{equation}\label{b3}
\begin{aligned}
g(k)[f(k)C_k]C_k+[g(k)C_k]f(k)C_k=[g(k)C_k][f(k)C_k],\\
\end{aligned}
\end{equation}
The formal solution of $\bar{\partial}$-problem (\ref{b1})
in terms of the matrix $R$ will be given from (\ref{b2}) as
\begin{equation}\label{b4}
\psi(k)=I\cdot(I-RC_k)^{-1}.
\end{equation}
For the sake of convenience, we define a pairing
$$\langle f,g\rangle=\frac{1}{2i\pi}\iint f(k)g^{\rm T}(k){\rm d}k\wedge{\rm d}\bar{k},\quad \langle f,g\rangle^{\rm T}=\langle g,f\rangle,$$
It is known that the above pairing possesses the following prosperities~\cite{BC}
\begin{equation}\label{b5}
\langle fR,g\rangle=\langle f,gR^{\rm T}\rangle,\quad \langle fC_k,g\rangle=-\langle f,gC_k\rangle.
\end{equation}
In addition, we can easily prove the following properties
\begin{equation}\label{b6}\begin{aligned}
&kf(k)C_k=k[f(k)C_k]+\langle f(k)\rangle,\\
&\frac{1}{\mu-k}f(k)C_k=\frac{1}{\mu-k}\{[f(k)C_k]-[f(\mu)C_\mu]\},
\end{aligned}
\end{equation}
where $\langle f(k)\rangle=\langle f(k),I\rangle$.

The aim of the $\bar{\partial}$ dressing method is to construct the compatible system of linear equations for $\psi$ and consequently the
nonlinear evolution equations associated the $\bar{\partial}$-problem (\ref{b1}). According to the main idea of the inverse scattering transform method,
it is important to introduce the $\xi,\tau$ dependence in the spectral transform matrix $R(k,\bar{k})$.
For the AB equations, let the $\xi$ and $\tau$-dependence be given by the linear and solvable equations
\begin{equation}\label{b7}
R_\xi=ik[\sigma_3,R],\quad \sigma_3={\rm diag}(1,-1),
\end{equation}
and
\begin{equation}\label{b14}
R_\tau=[\Omega,R],
\end{equation}
where $\Omega(k)$ is a singular dispersion relation, that is
\begin{equation}\label{b15}
\Omega(k)=\omega(k)C_k\sigma_3,
\end{equation}
where  $\omega(k)$ is some scalar function.
Differentiating (\ref{b2}) with respect to $\xi$ and $\tau$, and using (\ref{b7}),(\ref{b14}), as well as the properties of the Cauchy-Green operator (\ref{b6}),
we obtain the Zakharov-Shabat spectral problem~\cite{Z-G}
\begin{equation}\label{b13}
\begin{aligned}
&\psi_\xi-ik[\sigma_3,\psi]=Q\psi,\\
&\quad Q=i[\sigma_3,\langle\psi R\rangle],
\end{aligned}
\end{equation}
and the $\tau$-dependent linear equation associated with the singular dispersion relation
\begin{equation}\label{b18}
\psi_\tau=\left(\omega\psi\sigma_3\psi^{-1}C_k\right)\psi-\psi\Omega.
\end{equation}

\setcounter{equation}{0}
\section{\label{sec:three}The AB equations}
In this section, we will derive the AB equations equations associated with spectral problem (\ref{b13}).
To the end, differentiating the expression of $Q$ in (\ref{b13}) with respect to $\tau$ yields
\begin{equation}\label{c1}
Q_\tau=i[\sigma_3,\langle\psi R\rangle_\tau].
\end{equation}
Since $\bar\partial(f(k)C_k)=f(k)$, then
$$\begin{aligned}
(\psi R)_\tau&=\bar\partial\psi_\tau=\bar\partial\left\{\psi R_\tau C_k(I-RC_k)^{-1}\right\}\\
&=\bar\partial\left\{\psi R_\tau(I-RC_k)^{-1}C_k\right\}=\psi R_\tau(I-RC_k)^{-1}.
\end{aligned}$$
Hence, in virtue of the properties (\ref{b4}), equation (\ref{c1}) can be rewritten as
\begin{equation}\label{c2}
Q_\tau=i[\sigma_3,\langle\psi R_\tau(I-RC_k)^{-1},I\rangle]=i[\sigma_3,\langle\psi R_\tau,I\cdot(I+R^{\rm T}C_k)^{-1}\rangle].
\end{equation}
Based on the identity $\bar\partial(\psi^{-1})^{\rm T}=-(\psi^{-1})^{\rm T}R^{\rm T}$, the same procedure as (\ref{b2}) and (\ref{b4}) products
$$I\cdot(I+R^{\rm T}C_k)^{-1}=(\psi^{-1})^{\rm T}.$$
Therefore, using (\ref{b4}) and the definition of pairing $\langle f,g\rangle$, equation (\ref{c2}) takes the form
$$\begin{aligned}
Q_\tau&=i[\sigma_3,\langle\psi\Omega,(\psi^{-1}R^{\rm T})\rangle]-i[\sigma_3,\langle\psi R\Omega,(\psi^{-1})^{\rm T}\rangle]\\
&=-i[\sigma_3,\langle\psi\Omega,\bar\partial(\psi^{-1})^{\rm T}\rangle]-i[\sigma_3,\langle({\bar\partial}\psi)\Omega,(\psi^{-1})^{\rm T}\rangle]\\
&=-i[\sigma_3,\langle\psi\Omega\bar\partial\psi^{-1}\rangle]-i[\sigma_3,\langle({\bar\partial}\psi)\Omega\psi^{-1}\rangle].
\end{aligned}$$
Taking into account the fact that $\Omega\rightarrow0$ as $k\rightarrow\infty$, the above equation can be further reduced to
\begin{equation}\label{c3}\begin{aligned}
Q_\tau&=-i[\sigma_3,\langle\bar\partial(\psi\Omega\psi^{-1})\rangle-\langle\psi(\bar\partial\Omega)\psi^{-1}\rangle]\\
&=i[\sigma_3,\langle\omega(k)\psi \sigma_3\psi^{-1}\rangle].
\end{aligned}
\end{equation}

By virtue of the spectral problem (\ref{b13}), one can verify that
\begin{equation}\label{c4}
U_\xi=ik[\sigma_3,U]+[Q,U], \quad U=\psi \sigma_3\psi^{-1},
\end{equation}
and
\begin{equation}\label{c5}
Q_\tau=i[\sigma_3,\langle\omega(k)U\rangle].
\end{equation}
In order to derive the $\tau$-dependent linear spectral problem of the AB equations,
we take
\begin{equation}\label{c6}
\omega(k)=-i\pi\delta(k),
\end{equation}
then
\begin{equation}\label{c7}
V\equiv i\langle\omega U\rangle=-U|_{k=0},
\end{equation}
which implies
\begin{equation}\label{c8}
Q_\tau=[\sigma_3,V],\quad V_\xi=[Q,V].
\end{equation}
It is noted that the coupled equations (\ref{c8}) can also be derived from the compatibility condition of
the linear equations (\ref{b13}) and (\ref{b18}).

From (\ref{c6}), we know that the linear spectral problem (\ref{b18}) can be rewritten as
\begin{equation}\label{c9}
\psi_\tau+\frac{1}{ik}\psi\sigma_3=-\frac{1}{ik}V\psi.
\end{equation}

For the purpose of obtaining the AB equations, we introduce the following symmetry condition
\begin{equation}\label{c10}
Q^\dagger=-Q,
\end{equation}
from which we take
\begin{equation}\label{c11}
Q=2\left(\begin{matrix}
0&-\bar{A}\\
A&0\\
\end{matrix}\right).
\end{equation}
Here, the form of the potential function is chosen to ensure that the normalization condition $|A_\tau|^2+B^2=1$ can be obtained.
In addition, we need another symmetry condition about $\psi(k)$
\begin{equation}\label{c12}
\psi^\dagger(\bar{k})=\psi^{-1}(k).
\end{equation}
It is noted that this constraint condition can be obtained by using the symmetry condition (\ref{c10}) and
the spectral problem (\ref{b13}), as well as the linear equation (\ref{b7}).

From (\ref{c8}), we know that
$$\begin{aligned}
&V^{(o)}=\frac{1}{2}\sigma_3Q_\tau,\\
&Q_{\xi\tau}=[\sigma_3,[Q,V]]=2\sigma_3[Q,V^{(d)}],\\
&V^{(d)}_\xi=[Q,V^{(o)}]=-\frac{1}{2}\sigma_3(Q^2)_\tau,
\end{aligned}$$
where $V^{(o)}$ and $V^{(d)}$ denote the off-diagonal and diagonal of the matrix $V$, respectively.
Hence $V=V^{(o)}+V^{(d)}$. According to the above equations, we take
\begin{equation}\label{c13}
V=-\left(\begin{matrix}
B&\bar{A}_\tau\\
A_\tau&-B
\end{matrix}
\right),
\end{equation}
then we have the AB equations in canonical form
\begin{equation}\label{c14}
A_{\xi\tau}-4AB=0,\quad B_\xi+2(|A|^2)_\tau=0.
\end{equation}
It is remarked that the Lax pair of the AB equations is defined by (\ref{b13}) and (\ref{c9}), as well as (\ref{c13}).

\setcounter{equation}{0}
\section{\label{sec:four}Soliton solutions}
In the section, we will derive the explicit solutions of the AB equations (\ref{c14}) and their soliton solutions.
To this end, we introduce the spectral transform matrix $R$ as
\begin{equation}\label{d1}
R(k)=i\pi\sum\limits_{j=1}^N\left(\begin{matrix}
0&\bar{c}_je^{2ik\xi}\delta(k-\bar{k}_j)&\\
c_je^{-2ik\xi}\delta(k-k_j)&0\\
\end{matrix}\right),
\end{equation}
where $\{k_j\}_1^N$ are complex constants  and $c_j=c_j(\tau)$. 
The evolution of these $\tau$-dependent functions can be obtained
from (\ref{b14}) and (\ref{c6})
\begin{equation}\label{d2}
c_{j,\tau}=-\frac{2}{ik_j}c_j, \quad j=1,2,\cdots,N,
\end{equation}
Substituting (\ref{d1}) into (\ref{b13}),
in view of (\ref{c12}), yields
\begin{equation}\label{d3}
A=-i\langle\psi R\rangle_{21}=-\hat{\psi}_{22}\cdot g^T,
\end{equation}
where
\begin{equation}\label{d4}
\begin{aligned}
&\hat\psi_{22}=\left(\psi_{22}(k_1),\cdots,\psi_{22}(k_N)\right),\quad g=(g_1,\cdots,g_N),\\
&g_j=c_je^{2k_j\xi}=e^{2z_j}, \quad z_j=\theta_j-i\varphi_j,\\
&\quad\theta_j={\rm Im}k_j\xi+\frac{{\rm Im}k_j}{|k_j|^2}\tau+\kappa_j,\\
&\quad\varphi_j={\rm Re}k_j\xi-\frac{{\rm Re}k_j}{|k_j|^2}\tau+\chi_j,
\end{aligned}
\end{equation}
where $\{\kappa_j,\chi_j\}$ are arbitrary constants.
In addition, from (\ref{c7}) and (\ref{c4}),(\ref{c13}), by virtue of symmetry condition (\ref{c12}), we obtain
\begin{equation}\label{d5}
B=|\psi_{22}(0)|^2-|\psi_{21}(0)|^2,
\end{equation}
in terms of $\det\psi=1$.

In the following, we will give the expression of $\psi_{ij}$ about the discrete data.
Substitution (\ref{d1}) into (\ref{b2}) yields
\begin{equation}\label{d6}\begin{aligned}
\psi_{22}(k)&=1-i\sum\limits_{j=1}^N\frac{\psi_{21}(\bar{k}_j)\bar{g}_j}{\bar{k}_j-k},\\
\psi_{21}(k)&=-i\sum\limits_{j=1}^N\frac{\psi_{22}(k_j)g_j}{k_j-k},
\end{aligned}
\end{equation}
which imply that
\begin{equation}\label{d7}
\hat\psi_{22}=E(I+K\bar{K})^{-1},\quad\tilde\psi_{21}=-iEK(I+\bar{K}K)^{-1},
\end{equation}
where the vectors $\hat\psi_{22}, g$ are defined by (\ref{d4}) and
\begin{equation}\label{d8}\begin{aligned}
&\tilde\psi_{21}=(\psi_{21}(\bar{k}_1),\cdots,\psi_{21}(\bar{k}_N)),\quad E=(1,\cdots,1),\\
&K=(K_{nm})_{N\times N},\quad K_{nm}=\frac{g_n}{k_n-\bar{k}_m}.
\end{aligned}
\end{equation}
In addition, from (\ref{d6}), we have
\begin{equation}\label{d9}\begin{aligned}
&\psi_{21}(0)=-i\hat\psi_{22}h^T,\quad\psi_{22}(0)=1-i\tilde\psi_{21}\bar{h}^T,\\
&\quad h=(h_1,\cdots,h_N),\quad h_j=\frac{g_j}{k_j}.
\end{aligned}
\end{equation}

Substituting (\ref{d7}) into (\ref{d3}) and (\ref{d9})  one obtains \cite{H}
\begin{equation}\label{d10}
\begin{aligned}
A=&-{\rm tr}[(I+M)^{-1}g^TE]\\
=&-\frac{\det(I+M+g^TE)-\det(I+M)}{\det(I+M)},\\
\psi_{21}(0)=&-i{\rm tr}[(I+M)^{-1}h^TE]\\
=&-i\frac{\det(I+M+h^TE)-\det(I+M)}{\det(I+M)},\\
\psi_{22}(0)=&1-{\rm tr}[(I+\tilde{M})\bar{h}^TEK]\\
=&1-\frac{\det(I+\tilde{M}+\bar{h}^TEK)-\det(I+\tilde{M})}{\det(I+\tilde{M})},
\end{aligned}
\end{equation}
where
\begin{equation}\label{d11}
M=K\bar{K},\quad \tilde{M}=\bar{K}K.
\end{equation}
Indeed, for example, it is easy to see that $A=-E(I+M)^{-1}g^T$ in view of (\ref{d3}) and (\ref{d7}).
Then one may find $A=-{\rm tr}[(I+M)^{-1}g^TE]$ by multiplication of matrices, and
$A=-[\det(I+M+g^TE)-\det(I+M)]/\det(I+M)$ by the fact that $\det(g^TE)=0$.

In the following, we will give the one-soliton and two-soliton solutions. Firstly, for $N=1$,
$$\begin{aligned}
A=&-e^{-2i\varphi}\frac{2{\rm Im}k_1}{(2{\rm Im}k_1)e^{-2\theta_1}+e^{2\theta_1}(2{\rm Im}k)^{-1}},\\
\psi_{21}(0)=&\frac{e^{-2i\varphi}}{ik_1}\frac{2{\rm Im}k_1}{(2{\rm Im}k_1)e^{-2\theta_1}+e^{2\theta_1}(2{\rm Im}k)^{-1}},\\
\psi_{22}(0)=&1-\frac{e^{2\theta_1}}{i\bar{k}_1}\frac{1}{(2{\rm Im}k_1)e^{-2\theta_1}+e^{2\theta_1}(2{\rm Im}k)^{-1}},.
\end{aligned}$$
Let $2{\rm Im}k_1=e^{2\beta_1}$, then the above expressions give rise to one-soliton solution in semi-characteristic coordinates
\begin{equation}\label{d12}
\begin{aligned}
A=&-\frac{1}{2}e^{2(\beta_1-i\varphi_1)}{\rm sech}2(\theta_1-\beta_1),\\
B=&1+\frac{e^{2(\theta_1+\beta_1)}}{2|k_1|^2}[{\rm sech}^22(\theta_1-\beta_1)\sinh2(\theta_1-\beta_1)-{\rm sech}2(\theta_1-\beta_1)].
\end{aligned}
\end{equation}
It is readily verified that $\int_{-\infty}^\infty|A|^2d\xi={\rm Im}k_1/4$. One can find that
the waveform of the envelope solitary wave travels to the left, and the carrier wave to right, with same velocity $1/|k_1|^2$.
The graphic of one-soliton solution is shown in Figure 1.\\
\begin{figure}[h]
\centering
\includegraphics[width=5.0cm,angle=-90]{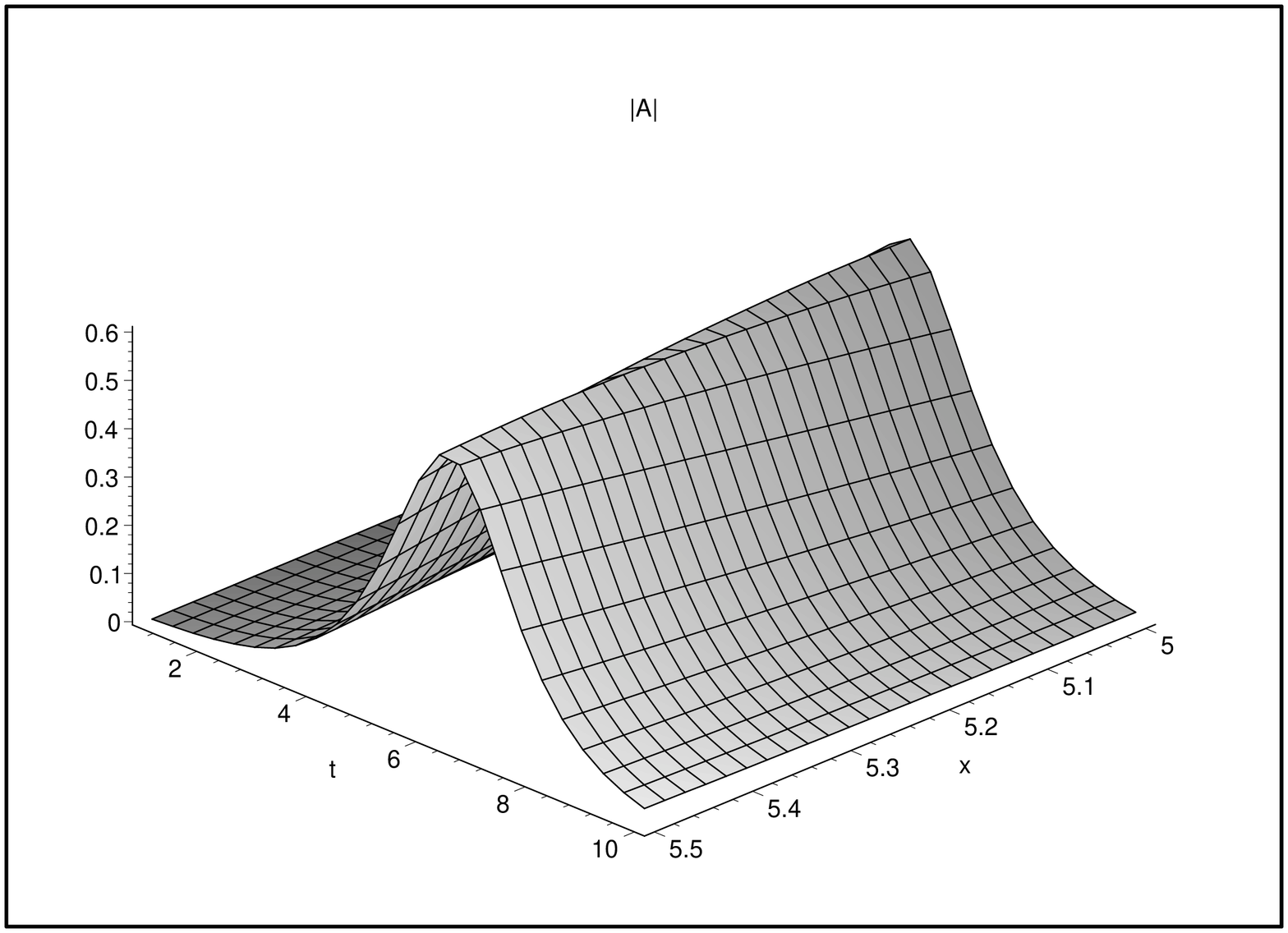}
\includegraphics[width=5.0cm,angle=-90]{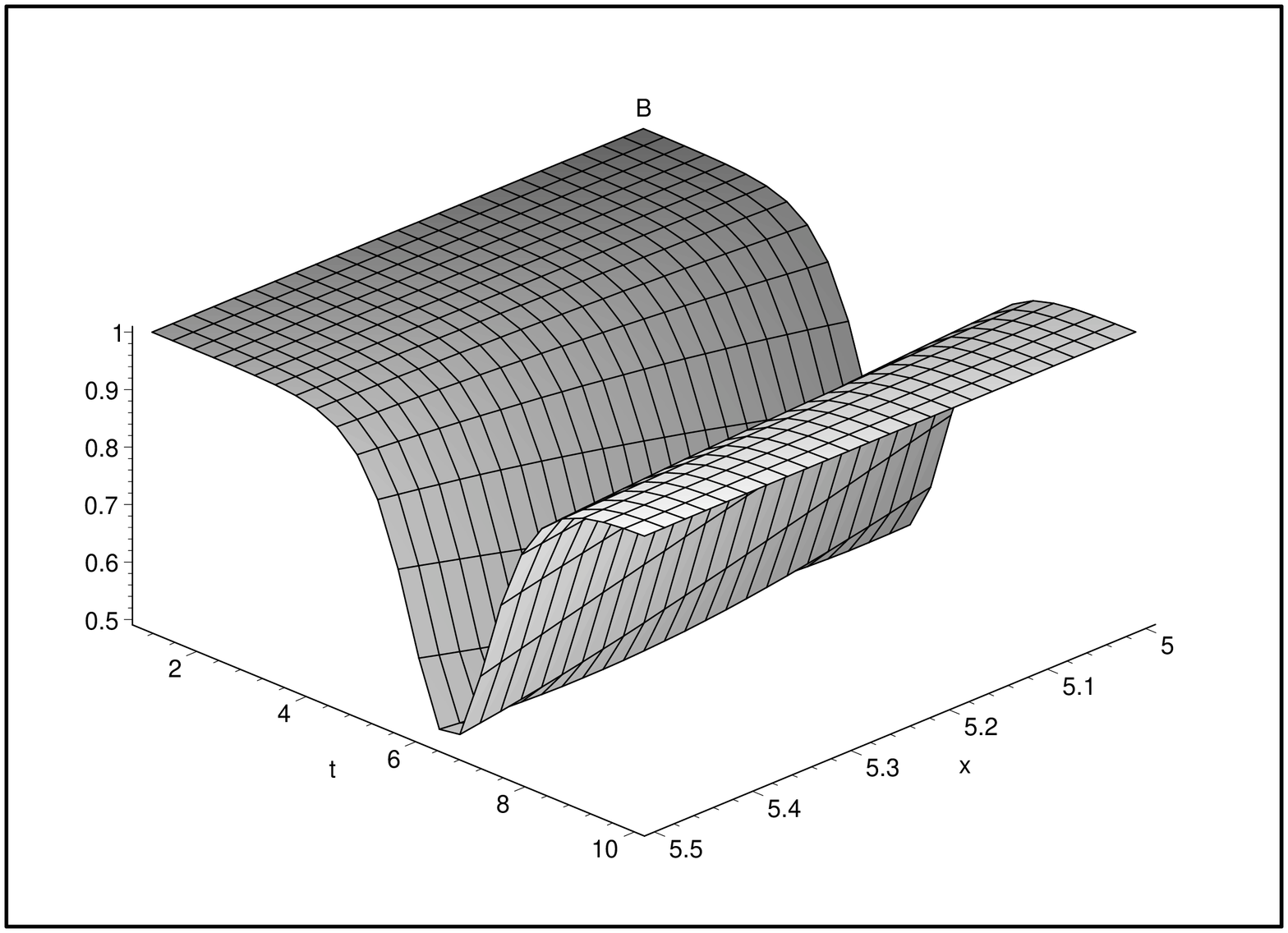}
\caption{$k_1=1.04+0.6i,\kappa_1=0,\xi_1=0$.}
\end{figure}

For the case of $N=2$, we have
\begin{equation}\label{d13}\begin{aligned}
\det(I+M)=&1+\frac{|k_1-k_2|^4}{\prod\limits_{j,l=1}^2(k_j-\bar{k}_l)^2}e^{2(z_1+z_2+\bar{z}_1+\bar{z}_2)}-\frac{e^{2(z_1+\bar{z}_1)}}{(k_1-\bar{k}_1)^2}\\
&-\frac{e^{2(z_2+\bar{z}_2)}}{(k_2-\bar{k}_2)^2}-\frac{e^{2(z_1+\bar{z}_2)}}{(k_1-\bar{k}_2)^2} -\frac{e^{2(\bar{z}_1+z_2)}}{(k_2-\bar{k}_1)^2},\\
\det(I+M+&g^TE)-\det(I+M)\\
=&e^{2z_1}\left[1-\frac{(k_1-k_2)^2}{(k_1-\bar{k}_2)^2(k_2-\bar{k}_2)^2}e^{2(z_2+\bar{z}_2)}\right]\\
&+e^{2z_2}\left[1-\frac{(k_1-k_2)^2}{(k_1-\bar{k}_1)^2(k_2-\bar{k}_1)^2}e^{2(z_1+\bar{z}_1)}\right],\\
\det(I+M+&h^TE)-\det(I+M)\\
=&\frac{e^{2z_1}}{k_1}\left[1-\frac{\bar{k}_2}{k_2}\frac{(k_1-k_2)^2}{(k_1-\bar{k}_2)^2(k_2-\bar{k}_2)^2}e^{2(z_2+\bar{z}_2)}\right]\\
&+\frac{e^{2z_2}}{k_2}\left[1-\frac{\bar{k}_1}{k_1}\frac{(k_1-k_2)^2}{(k_1-\bar{k}_1)^2(k_2-\bar{k}_1)^2}e^{2(z_1+\bar{z}_1)}\right],\\
\end{aligned}
\end{equation}
and
$$\begin{aligned}
2\det(I+\tilde{M})&-\det(I+\tilde{M}+\bar{h}^TEK)\\
=&1-\frac{k_1}{\bar{k}_1}\frac{e^{2(z_1+\bar{z}_1)}}{(k_1-\bar{k}_1)^2}-\frac{k_2}{\bar{k}_2}\frac{e^{2(z_2+\bar{z}_2)}}{(k_2-\bar{k}_2)^2}
-\frac{k_2}{\bar{k}_1}\frac{e^{2(\bar{z}_1+z_2)}}{(k_2-\bar{k}_1)^2}\\
&-\frac{k_1}{\bar{k}_2}\frac{e^{2(z_1+\bar{z}_2)}}{(k_1-\bar{k}_2)^2}+\frac{k_1k_2}{\bar{k}_1\bar{k}_2}\frac{|k_1-k_2|^4}{\prod\limits_{j,l=1}^2(k_j-\bar{k}_l)^2}e^{2(z_1+z_2+\bar{z}_1+\bar{z}_2)}.
\end{aligned}$$
To obtain the soliton solutions, we introduce new functions $\omega_j$
\begin{equation}\label{d14}
e^{2\omega_j}=\frac{k_1-k_2}{(k_1-\bar{k}_j)(k_2-\bar{k}_j)},\quad \omega_j=w_j+i\phi_j,\quad j=1,2.
\end{equation}
Under this definition, equations (\ref{d13}) can be rewritten as
\begin{equation}\label{d15}
\begin{aligned}
\det(I+M)
=&4\frac{e^{2(\vartheta_1+\vartheta_2)}}{|k_1-k_2|^2}\left\{a\cosh2\vartheta_1\cosh2\vartheta_2+b\sinh2\vartheta_1\sinh2\vartheta_2\right.\\
&\quad\left.+2{\rm Im}k_1{\rm Im}k_2\cos\rho \right\}\\
\equiv&\Delta D_2,
\end{aligned}
\end{equation}
\begin{equation}\label{d16}
\begin{aligned}
\det(I+M&+g^TE)-\det(I+M)\\
=&-4\frac{e^{2(\vartheta_1+\vartheta_2)}}{|k_1-k_2|^2}\sqrt{a^2-b^2}\left[{\rm Im}k_1e^{2i(\phi_2-\varphi_1)}\sinh2(\vartheta_2+i\phi_2)\right.\\
&\qquad+\left.{\rm Im}k_2e^{2i(\phi_1-\varphi_2)}\sinh2(\vartheta_1+i\phi_1)\right]\\
&\equiv\Delta \Omega_2,
\end{aligned}
\end{equation}
\begin{equation}\label{d17}
\begin{aligned}
\det(I+M&+h^TE)-\det(I+M)\\
=&-4\frac{e^{2(\vartheta_1+\vartheta_2)}}{|k_1-k_2|^2}\sqrt{a^2-b^2}\left[\frac{{\rm Im}k_1}{k_1}e^{2i(\tilde{\phi}_2-\varphi_1)}\sinh2(\vartheta_2+i\tilde{\phi}_2)\right.\\
&\qquad+\left.\frac{{\rm Im}k_2}{k_2}e^{2i(\tilde{\phi_1}-\varphi_2)}\sinh2(\vartheta_1+i\tilde{\phi}_1)\right]\\
&\equiv\Delta \Xi_2,
\end{aligned}
\end{equation}
\begin{equation}\label{d18}
\begin{aligned}
&2\det(I+\tilde{M})-\det(I+\tilde{M}+\bar{h}^TgK)\\
&=4\frac{e^{2(\vartheta_1+\vartheta_2)}}{|k_1-k_2|^2}\left\{e^{i(\arg k_1+\arg k_2)}[a\cosh(2\vartheta_1+i\arg k_1)\cosh(2\vartheta_2+i\arg k_2)\right.\\
&\qquad\left.b\sinh(2\vartheta_1+i\arg k_1)\sinh(2\vartheta_2+i\arg k_2)]
+2{\rm Im}k_1{\rm Im}k_2\cosh(\varepsilon+i\varrho)\right\}\\
&\quad\equiv\Delta\Lambda_2,
\end{aligned}
\end{equation}
where
$$\begin{aligned}
&\vartheta_j=\theta_j+w_j,\ \tilde\phi_j=\phi_j+\varpi_j,\ \varpi_j=\arg k_j, j=1,2\\
& |k_1-k_2|^2=a+b,\ |k_1-\bar{k}_2|^2=a-b,\\
&\ \rho=2(\varphi_1+\phi_1-\varphi_2-\phi_2),\\
&\varrho=2(\varphi_1-\phi_1-\varphi_2+\phi_2)-\varpi_1+\varpi_2,\\
&e^\varepsilon=\frac{|k_1|}{|k_2|},\quad\Delta=4\frac{e^{2(\vartheta_1+\vartheta_2)}}{|k_1-k_2|^2}\cosh2\vartheta_1\cosh2\vartheta_2,
\end{aligned}$$
and
\begin{equation}\label{d19}\begin{aligned}
D_2=&a+b\tanh2\vartheta_1\tanh2\vartheta_2+2{\rm Im}k_1{\rm Im}k_2\cos\rho{\rm sech}2\vartheta_1{\rm sech}2\vartheta_2,\\
\Omega_2=&-\sqrt{a^2-b^2}[{\rm Im}k_1e^{2i(\phi_2-\varphi_1)}{\rm sech}2\vartheta_1(\tanh2\vartheta_2\cos2\phi_2+i\sin2\phi_2)\\
&\quad+{\rm Im}k_2e^{2i(\phi_1-\varphi_2)}{\rm sech}2\vartheta_2(\tanh2\vartheta_1\cos2\phi_1+i\sin2\phi_1)],\\
\Xi_2=&-\sqrt{a^2-b^2}[\frac{{\rm Im}k_1}{k_1}e^{2i(\tilde{\phi}_2-\varphi_1)}{\rm sech}2\vartheta_1(\tanh2\vartheta_2\cos2\tilde{\phi}_2+i\sin2\tilde{\phi}_2)\\
&\quad+\frac{{\rm Im}k_2}{k_2}e^{2i(\tilde{\phi}_1-\varphi_2)}{\rm sech}2\vartheta_2(\tanh2\vartheta_1\cos2\tilde{\phi}_1+i\sin2\tilde{\phi}_1)],\\
\Lambda_2=&e^{i(\varpi_1+\varpi_2)}[a(\cos\varpi_1\cos\varpi_2-\tanh2\vartheta_1\tanh2\vartheta_2\sin\varpi_1\sin\varpi_2)\\
&\quad+b(\tanh2\vartheta_1\tanh2\vartheta_2\cos\varpi_1\cos\varpi_2-\sin\varpi_1\sin\varpi_2)\\
&\quad+ia(\tanh2\vartheta_2\cos\varpi_1\sin\varpi_2+\tanh2\vartheta_1\sin\varpi_1\cos\varpi_2)\\
&\quad+ib(\tanh2\vartheta_1\cos\varpi_1\sin\varpi_2+\tanh2\vartheta_2\sin\varpi_1\cos\varpi_2)]\\
&\qquad+2{\rm Im}k_1{\rm Im}k_2\cosh(\varepsilon+i\varrho){\rm sech}2\vartheta_1{\rm sech}2\vartheta_2.
\end{aligned}
\end{equation}

Hence, the two-soliton solution in semi-characteristic coordinates of the AB equations (\ref{c14}) takes the form
\begin{equation}\label{d20}
A=\frac{\Omega_2}{D_2},\quad B=\frac{|\Lambda_2|^2-|\Xi_2|^2}{D_2^2}.
\end{equation}
The Figure 2 describes two-soliton waves of $|A|$ and $B$ traveling to left, and Figure 3 (from left to right) shows collision of the two-soliton from $|A|$ at
$\tau_1=-3,\tau_2=0,\tau_3=3$; Figure 4 (from left to right) shows collision of the two-soliton from $B$ at
$\tau_1=-4,\tau_2=-1,\tau_3=2$. (In the figures, for convenience, we take variable $\xi$ as $x$, and $\tau$ as $t$.)
\begin{figure}[h]
\centering
\includegraphics[width=5.0cm,angle=-90]{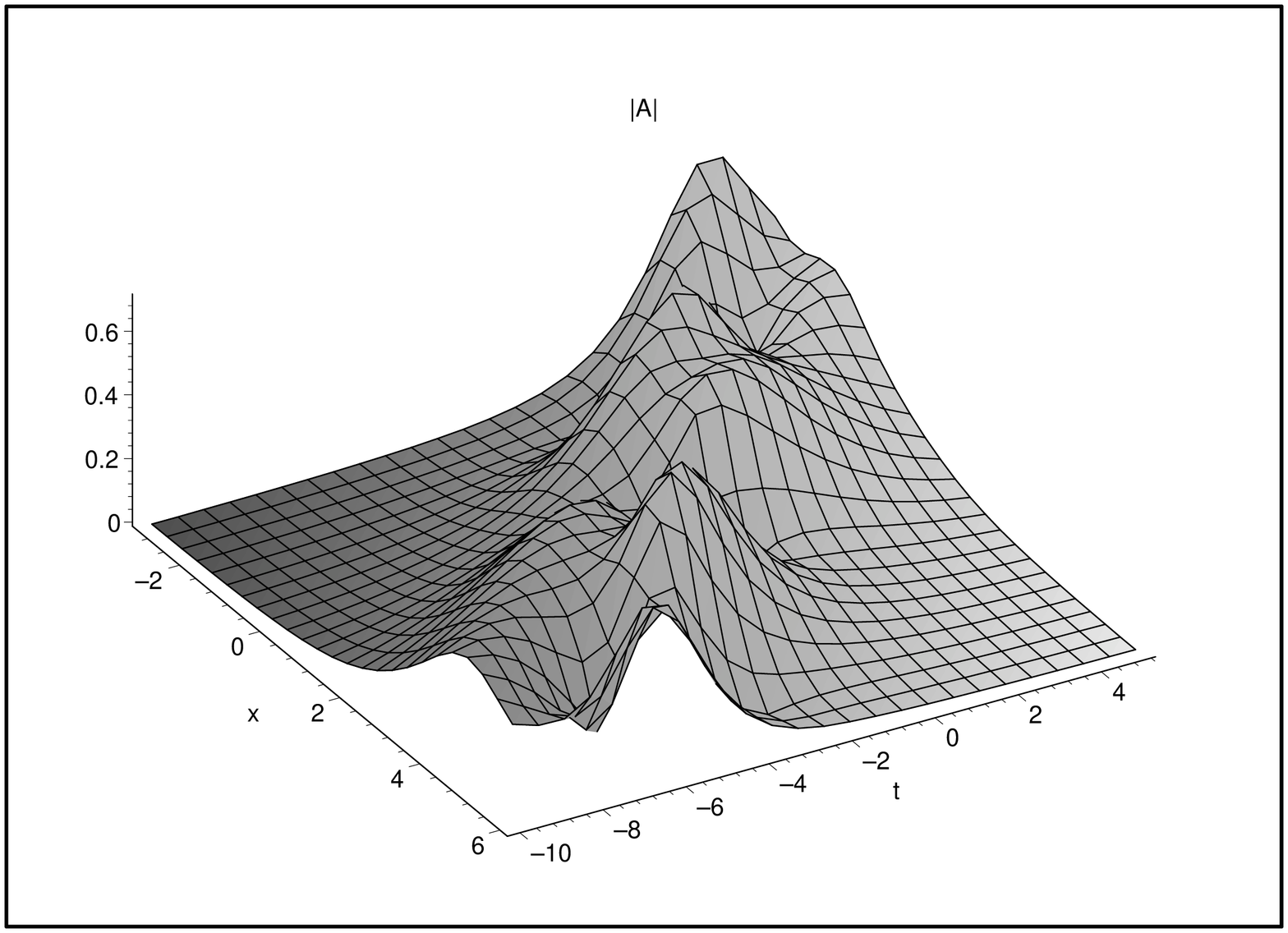}
\includegraphics[width=5.0cm,angle=-90]{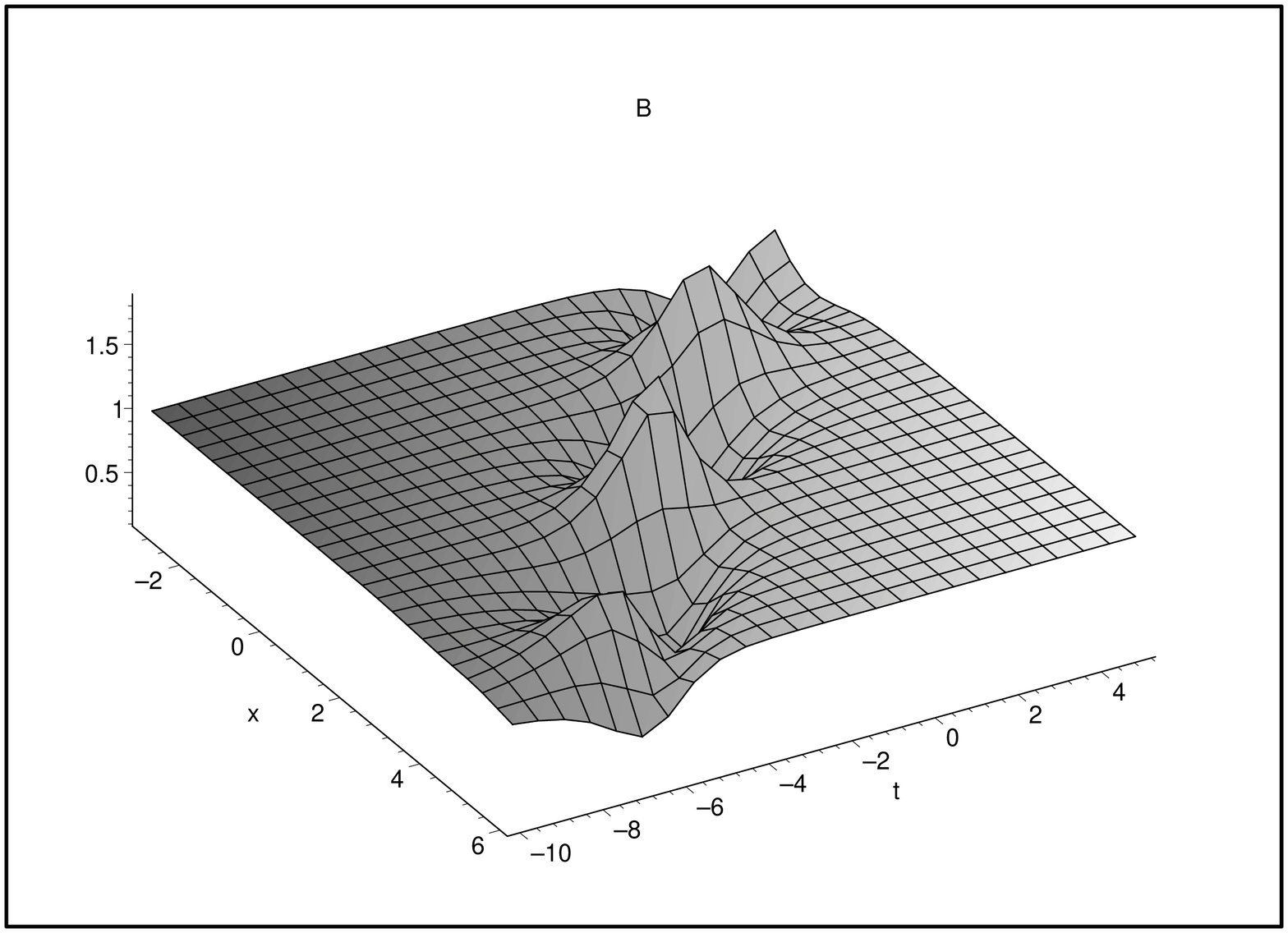}
\caption{$k_1=1.04+0.6i,k_2=2+0.4\mathrm{i},\kappa_j=0,\xi_j=0,j=1,2$.}
\end{figure}
\vskip9cm
\begin{figure}[h]
\centering
\includegraphics[width=3.0cm,angle=-90]{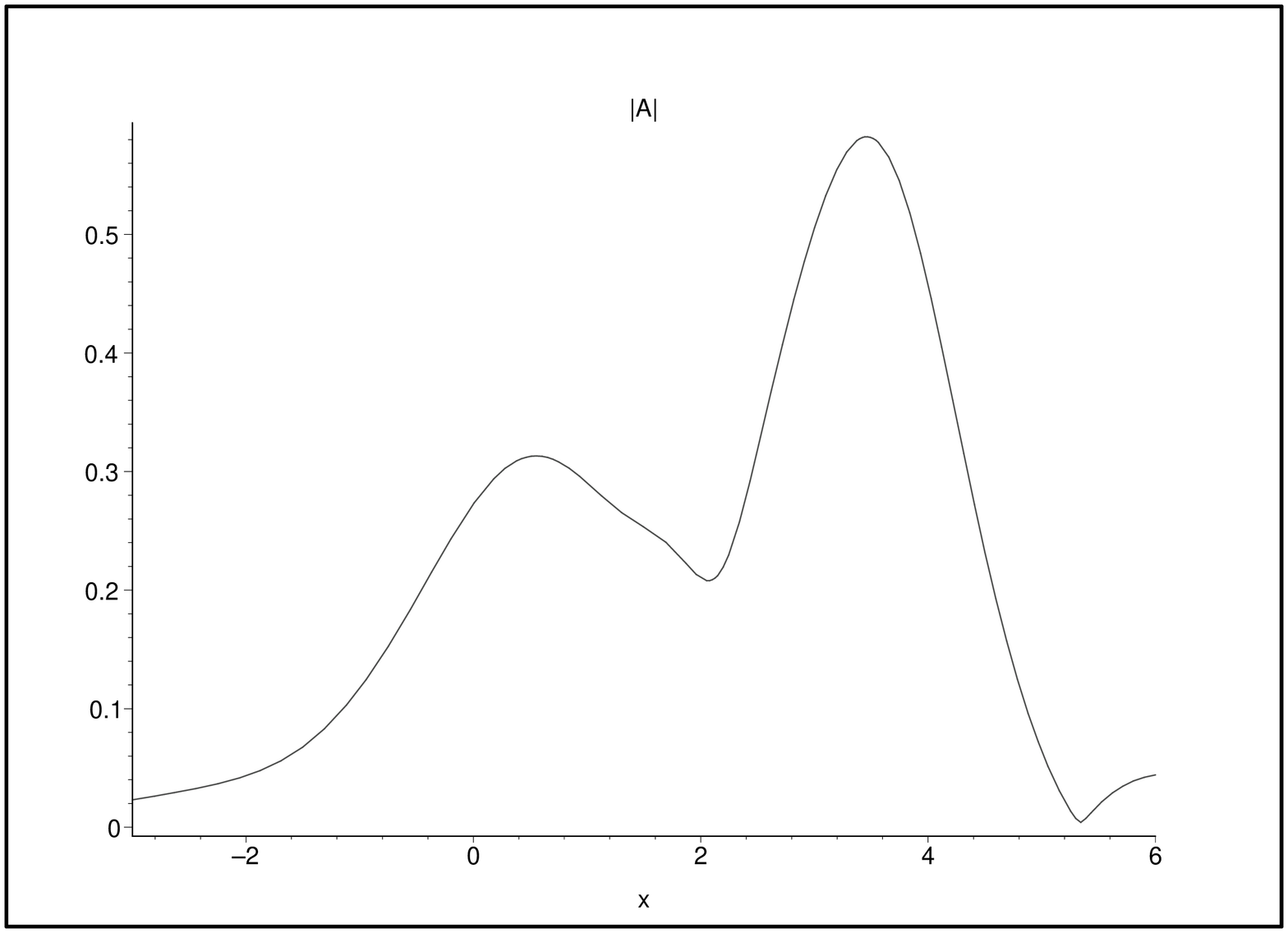}
\includegraphics[width=3.0cm,angle=-90]{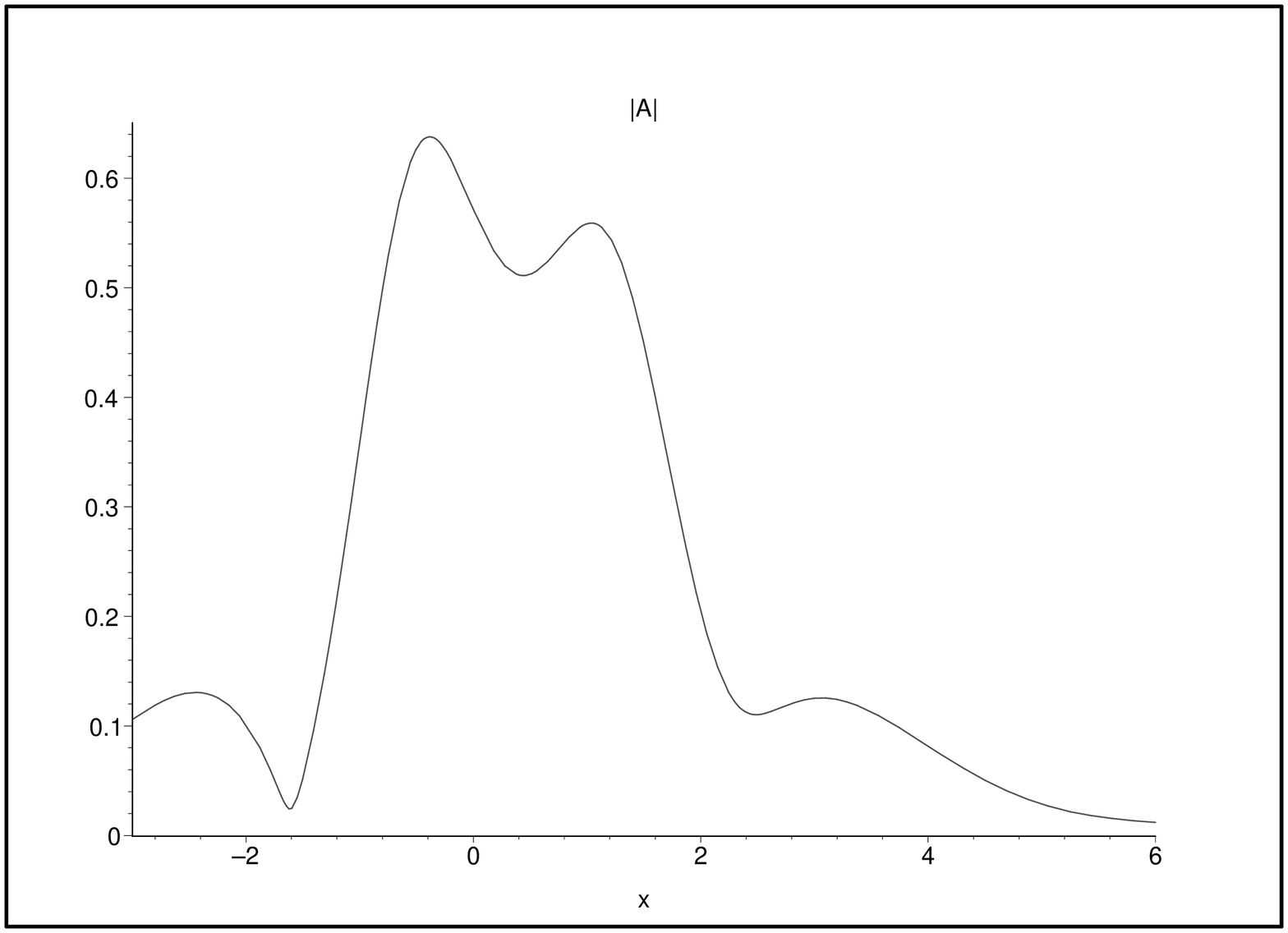}
\includegraphics[width=3.0cm,angle=-90]{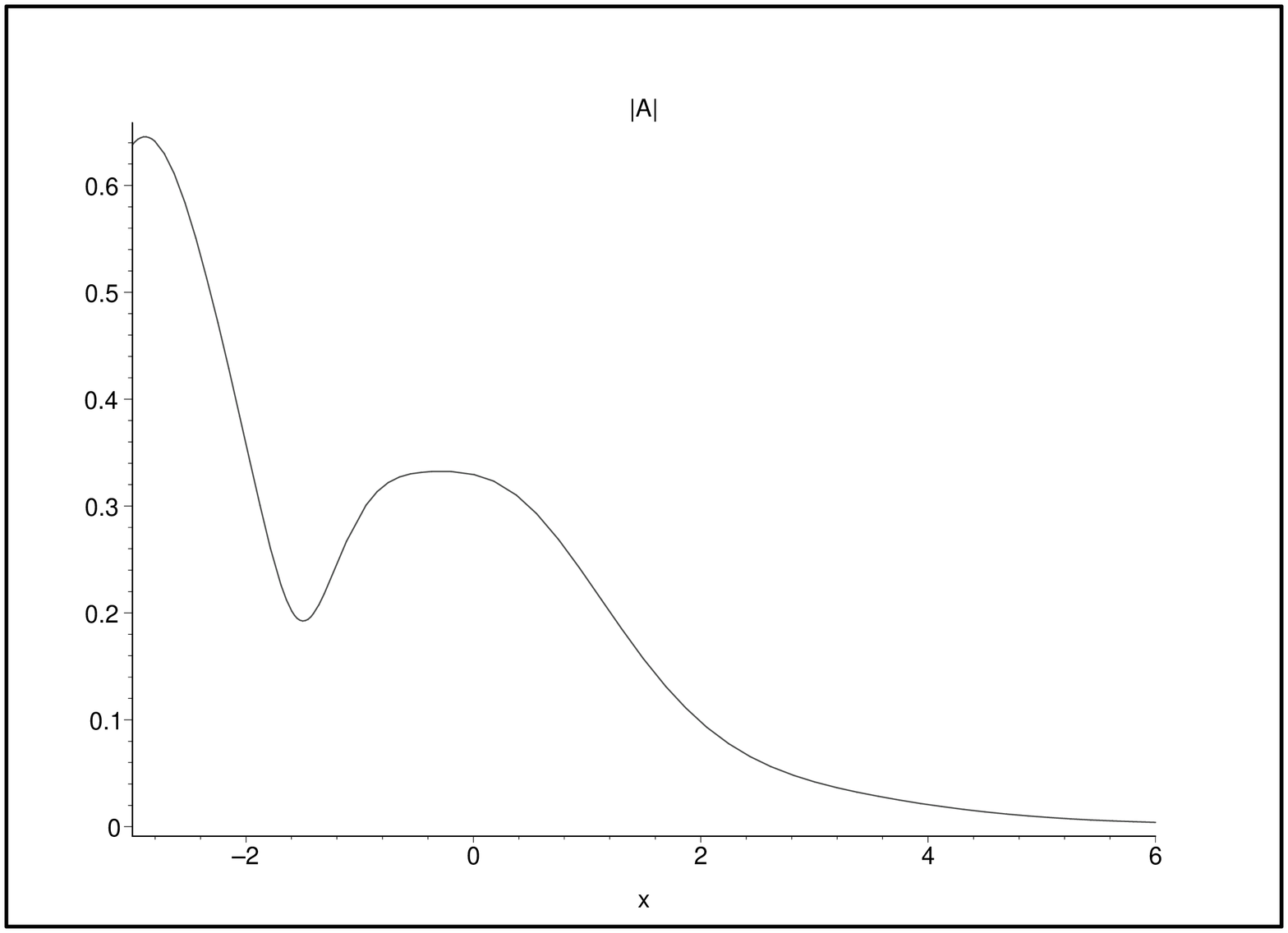}
\caption{$k_1=1.04+0.6i,k_2=2+0.4\mathrm{i},\kappa_j=0,\xi_j=0,j=1,2$.}
\end{figure}
\begin{figure}[h]
\centering
\includegraphics[width=3.0cm,angle=-90]{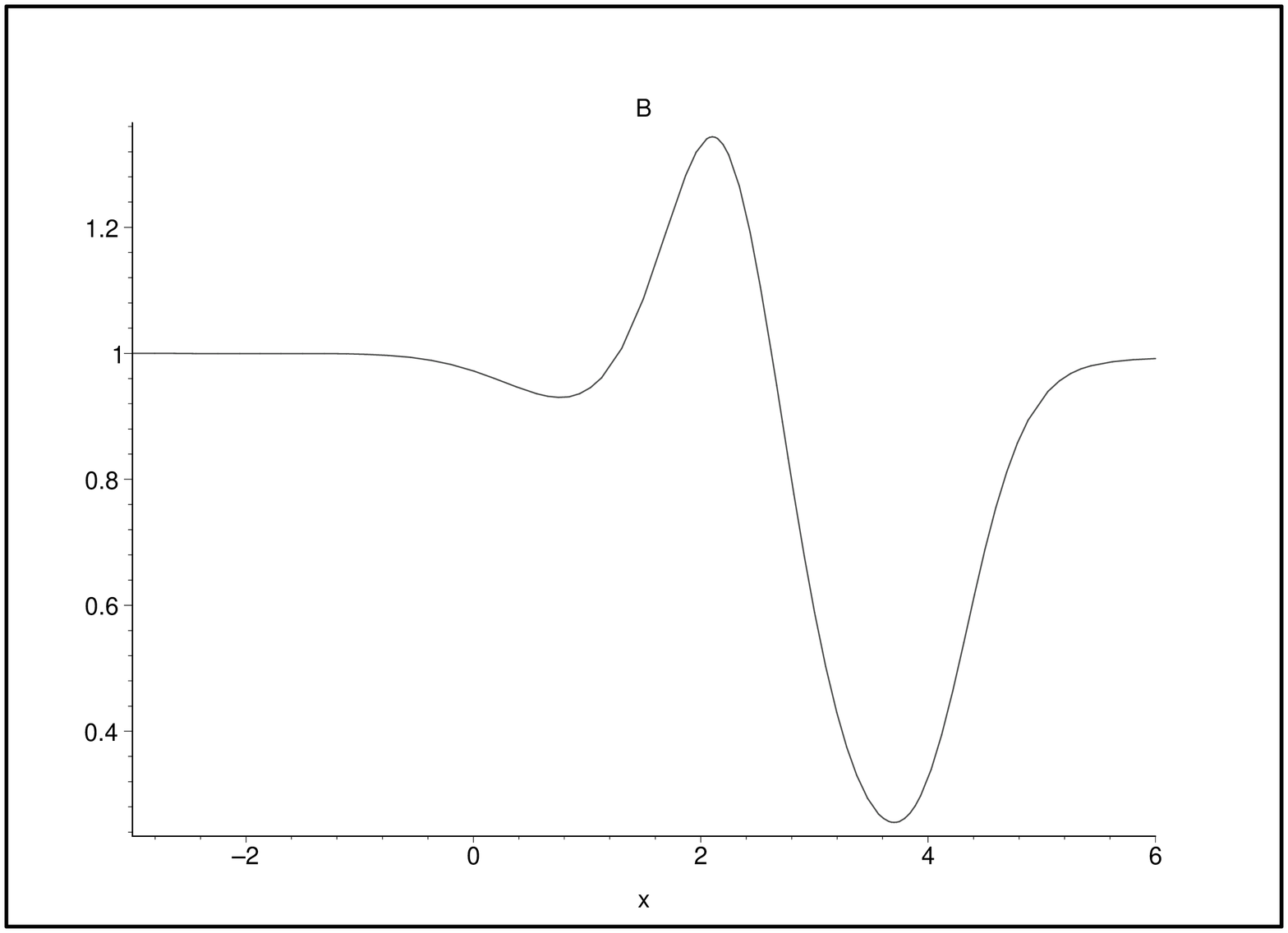}
\includegraphics[width=3.0cm,angle=-90]{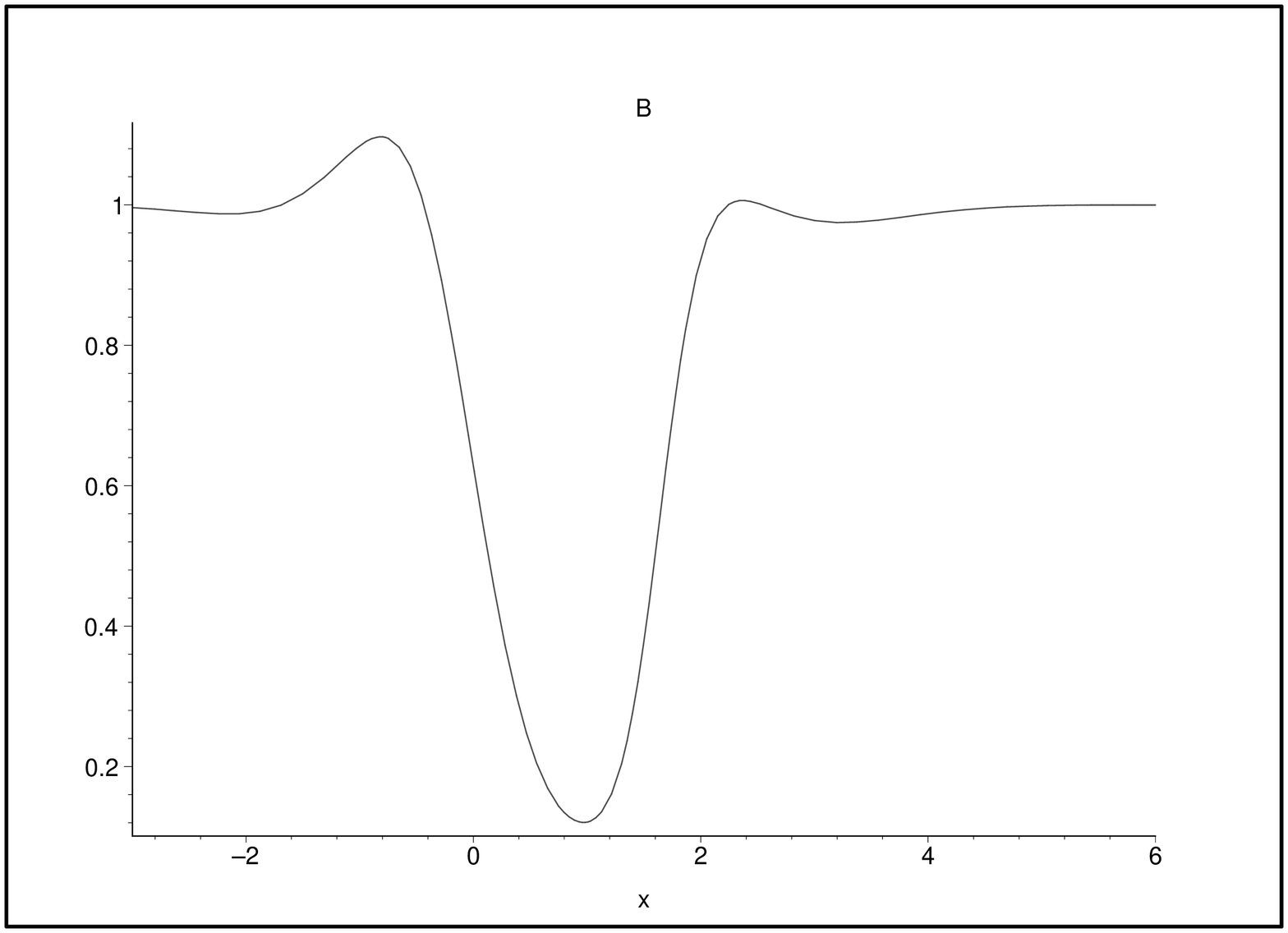}
\includegraphics[width=3.0cm,angle=-90]{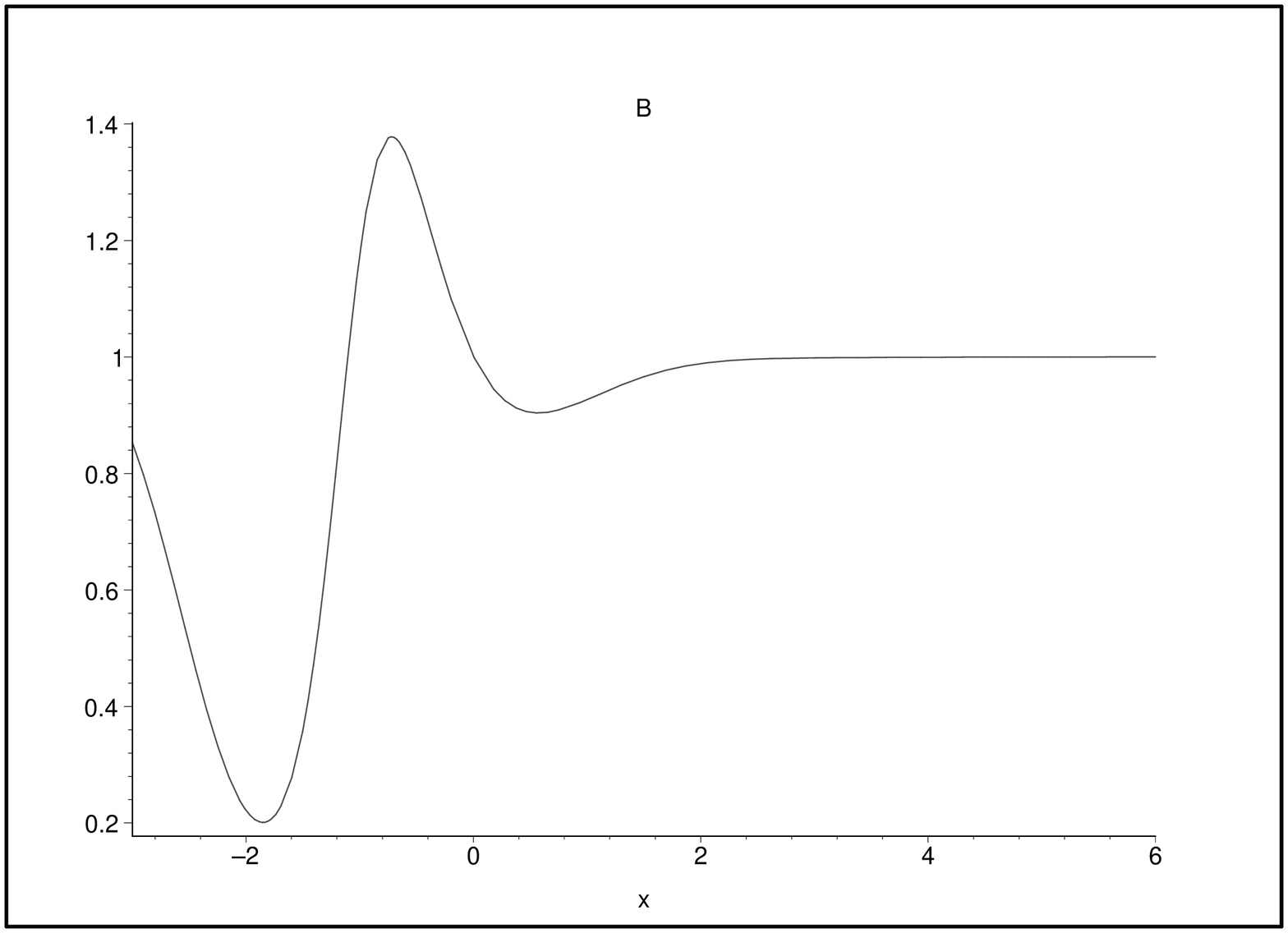}
\caption{$k_1=1.04+0.6i,k_2=2+0.4\mathrm{i},\kappa_j=0,\xi_j=0,j=1,2$.}
\end{figure}
Note that the representations (\ref{d15})-(\ref{d18}) can be rewritten into another forms
\begin{equation}\label{d21}
\begin{aligned}
&\det(I+M)=\tilde\Delta\tilde{D_2},\quad \det(I+M+g^TE)-\det(I+M)=\tilde\Delta\tilde\Omega_2\\
&\det(I+M+h^TE)-\det(I+M)=\tilde\Delta\tilde\Xi_2,\\
&2\det(I+\tilde{M})-\det(I+\tilde{M}+\bar{h}^TgK)=\tilde\Delta\tilde\Lambda_2,
\end{aligned}
\end{equation}
where
$$\tilde\Delta=4\frac{e^{2(\vartheta_1+\vartheta_2)}}{|k_1-k_2|^2}\sinh2\vartheta_1\sinh2\vartheta_2,$$
and
\begin{equation}\label{d22}\begin{aligned}
\tilde{D_2}=&b+a\coth2\vartheta_1\coth2\vartheta_2+2{\rm Im}k_1{\rm Im}k_2\cos\rho{\rm csch}2\vartheta_1{\rm csch}2\vartheta_2,,\\
\tilde\Omega_2=&-\sqrt{a^2-b^2}[{\rm Im}k_1e^{2i(\phi_2-\varphi_1)}{\rm csch}2\vartheta_1(\cos2\phi_2+i\coth2\vartheta_2\sin2\phi_2)\\
&\quad+{\rm Im}k_2e^{2i(\phi_1-\varphi_2)}{\rm csch}2\vartheta_2(\cos2\phi_1+i\coth2\vartheta_1\sin2\phi_1)],\\
\tilde\Xi_2=&-\sqrt{a^2-b^2}[\frac{{\rm Im}k_1}{k_1}e^{2i(\tilde{\phi}_2-\varphi_1)}{\rm csch}2\vartheta_1(\cos2\tilde{\phi}_2+i\coth2\vartheta_2\sin2\tilde{\phi}_2)\\
&\quad+\frac{{\rm Im}k_2}{k_2}e^{2i(\tilde{\phi}_1-\varphi_2)}{\rm csch}2\vartheta_2(\cos2\tilde{\phi}_1+i\coth2\vartheta_1\sin2\tilde{\phi}_1)],\\
\tilde\Lambda_2=&e^{i(\varpi_1+\varpi_2)}[a(\coth2\vartheta_1\coth2\vartheta_2\cos\varpi_1\cos\varpi_2-\sin\varpi_1\sin\varpi_2)\\
&\quad+b(\cos\varpi_1\cos\varpi_2-\coth2\vartheta_1\coth2\vartheta_2\sin\varpi_1\sin\varpi_2)\\
&\quad+ia(\coth2\vartheta_1\cos\varpi_1\sin\varpi_2+\coth2\vartheta_2\sin\varpi_1\cos\varpi_2)\\
&\quad+ib(\coth2\vartheta_2\cos\varpi_1\sin\varpi_2+\coth2\vartheta_1\sin\varpi_1\cos\varpi_2)]\\
&\qquad+2{\rm Im}k_1{\rm Im}k_2\cosh(\varepsilon+i\varrho){\rm csch}2\vartheta_1{\rm csch}2\vartheta_2.
\end{aligned}
\end{equation}
Hence, we have another form of two-soliton solution of the AB equations (\ref{c14})
\begin{equation}\label{d23}
A=\frac{\tilde\Omega_2}{\tilde{D_2}},\quad B=\frac{|\tilde\Lambda_2|^2-|\tilde\Xi_2|^2}{\tilde{D_2}^2}.
\end{equation}
It is remarked that the functions $\vartheta_j$ can be rewritten as
\begin{equation}\label{d24}
\vartheta_j={\rm Im}k_j(\xi+\frac{\tau}{|k_j|^2}+\xi_j),
\end{equation}
where $\xi_j$ is a certain constant. We note that the denominators $D_2$ and $\tilde{D}_2$ are not zero, because they are derived
from the determinant of an invertible matrix. It is remarked that the graphic of solution (\ref{d23}) has the same form as in Figure 2.

Now, we will derive the $N$-soliton solutions of (\ref{c14}).
By virtue of the method of linear algebra, we know that
\begin{equation}\label{d25}
\det(I+M)=1+\sum\limits_{\sigma=1}^N\sum\limits_{1\leq j_1\leq\cdots\leq j_\sigma\leq N}M(j_1,\cdots,j_\sigma),
\end{equation}
where $M(j_1,\cdots,j_\sigma)$ denotes the principal minor of $N\times N$ matrix $M$ obtained
by taking all the elements of $(j_1,\cdots,j_\sigma)$-th columns and rows.
By using the Cauchy-Binet formula,  we can calculate the value of $M(j_1,\cdots,j_\sigma)$
\begin{equation}\label{d26}
M(j_1,\cdots,j_\sigma)=\sum\limits_{1\leq r_1\leq\cdots\leq r_\sigma\leq N}K\left(\begin{aligned}
j_1,&j_2,&\cdots,&j_\sigma\\
r_1,&r_2,&\cdots,&r_\sigma
\end{aligned}\right)\bar{K}\left(\begin{aligned}
r_1,&r_2,&\cdots,&r_\sigma\\
j_1,&j_2,&\cdots,&j_\sigma
\end{aligned}\right),
\end{equation}
where $K\left(\begin{aligned}
j_1,&j_2,&\cdots,&j_\sigma\\
r_1,&r_2,&\cdots,&r_\sigma
\end{aligned}\right)$ denotes the determinant of the submatrix obtained by preserving the $(j_1,j_2,\cdots,j_\sigma)$-th rows
and $(r_1,r_2,\cdots,r_\sigma)$-th columns of $K$; $\bar{K}\left(\begin{aligned}
\cdot\\
\cdot
\end{aligned}\right)$ denotes similarly the determinant of the submatrix for $\bar{K}$.
It is noted that $K$ is Cauchy type matrices, then
\begin{equation}\label{d27}
M(j_1,\cdots,j_\sigma)=\sum\limits_{1\leq r_1\leq\cdots\leq r_\sigma\leq N}(-1)^\sigma
\prod\limits_{l<l^\prime,m<m^\prime}\frac{|k_l-k_{l^\prime}|^2|\bar{k}_{m^\prime}-\bar{k}_m|^2}{(k_l-\bar{k}_m)^2}e^{2(z_l+\bar{z}_m)},
\end{equation}
where $m\in\{r_1,r_2,\cdots,r_\sigma\};l,l^\prime\in\{j_1,j_2,\cdots,j_\sigma\}$ and $\sigma=1,\cdots,N$.
Hence, we obtain the explicit representation of $\det(I+M)$
from (\ref{d25}) and (\ref{d27}). It is readily verified that $\det(I+\tilde{M})=\det(I+M)$.

In the following, we will evaluate the numerator of the expressions in (\ref{d10}). To this end, let
\begin{equation}\label{d28}
C=M+g^TE=GH,
\end{equation}
where $G=(g^T,K)=(G_{nm})$ and $H=\left(\begin{array}{c}
E\\
\bar{K}
\end{array}\right)=(H_{mn})$, with $n\in\{1,\cdots,N\}, m\in\{0,1,\cdots,N\}$.
Hence, $\det(I+C)$ takes the same expansion as (\ref{d25}), where
\begin{equation}\label{d29}
C(j_1,\cdots,j_\sigma)=\sum\limits_{0\leq r_1\leq\cdots\leq r_\sigma\leq N}G\left(\begin{aligned}
j_1,&j_2,&\cdots,&j_\sigma\\
r_1,&r_2,&\cdots,&r_\sigma
\end{aligned}\right)H\left(\begin{aligned}
r_1,&r_2,&\cdots,&r_\sigma\\
j_1,&j_2,&\cdots,&j_\sigma
\end{aligned}\right).
\end{equation}
Now, we split the summation on the right hand side of the above equation into two parts, the first one is $r_1=0$,
and the second one is $r_1\geq1$. It is noted that the second one is exactly equal to $M(j_1,\cdots,j_\sigma)$. Thus, the
numerator of the expression of $r$ in (\ref{d10}) takes the value 
$$\begin{aligned}
&\det(I+M+g^TE)-\det(I+M)\\
&=\sum\limits_{\sigma=1}^N\sum\limits_{(1\leq j_1\leq\cdots\leq j_\sigma\leq N)}\sum\limits_{(1\leq r_2\leq\cdots\leq r_\sigma\leq N)}
G\left(\begin{aligned}
j_1,&j_2,&\cdots,&j_\sigma\\
0,&r_2,&\cdots,&r_\sigma
\end{aligned}\right)H\left(\begin{aligned}
0,&r_2,&\cdots,&r_\sigma\\
j_1,&j_2,&\cdots,&j_\sigma
\end{aligned}\right)\\
&=\sum\limits_{\sigma=1}^N\sum\limits_{(1\leq j_1\leq\cdots\leq j_\sigma\leq N)}\sum\limits_{(1\leq r_2\leq\cdots\leq r_\sigma\leq N)}
(-1)^{\sigma-1}\prod\limits_{l<l^\prime,m<m^\prime}\frac{|k_l-k_{l^\prime}|^2|\bar{k}_{m^\prime}-\bar{k}_m|^2}{(k_l-\bar{k}_m)^2}e^{2(z_l+\bar{z}_m)},
\end{aligned}$$
where $m,m^\prime\in\{r_2,\cdots,r_\sigma\};   l,l^\prime\in\{j_1,j_2,\cdots,j_\sigma\}.$

Similarly, for $\psi_{21}(0)$ in (\ref{d10}), we have
\begin{equation}\label{d30}
\begin{aligned}
\det(I&+M+h^TE)-\det(I+M)\\
&=\sum\limits_{\sigma=1}^N\sum\limits_{(1\leq j_1\leq\cdots\leq j_\sigma\leq N)}\sum\limits_{(1\leq r_2\leq\cdots\leq r_\sigma\leq N)}\\
&\quad\times(-1)^{\sigma-1}\prod\limits_{l<l^\prime,m<m^\prime}\frac{\bar{k}_m}{k_l}\frac{|k_l-k_{l^\prime}|^2|\bar{k}_{m^\prime}-\bar{k}_m|^2}{(k_l-\bar{k}_m)^2}e^{2(z_l+\bar{z}_m)},\\
\end{aligned}
\end{equation}
where $(m,m^\prime\in\{r_2,\cdots,r_\sigma\}; l,l^\prime\in\{j_1,j_2,\cdots,j_\sigma\}).$
While for $\psi_{22}(0)$, let
$$\begin{aligned}
\tilde{M}+\bar{h}^TEK=(\bar{h}^T,\bar{K})\left(\begin{array}{c}
EK\\
K
\end{array}\right),
\end{aligned}$$
then
\begin{equation}\label{d31}\begin{aligned}
&\det(I+\tilde{M}+\bar{h}^TgK)-\det(I+\tilde{M})\\
&=\sum\limits_{\sigma=1}^N\sum\limits_{(1\leq j_1\leq\cdots\leq j_\sigma\leq N)}\sum\limits_{(1\leq r_2\leq\cdots\leq r_\sigma\leq N)}
(-1)^{\sigma-1}\sum\limits_{r_0\in\hat\sigma}\prod\limits_{l,m}\frac{k_{r_0}-k_m}{k_{r_0}-\bar{k}_l}e^{2z_{r_0}}\\
&\qquad\times\prod\limits_{l,m}\frac{k_m}{\bar{k}_l}\frac{e^{2(\bar{z}_m+z_l)}}{(k_m-\bar{k}_l)^2}\prod\limits_{l<l^\prime,m<m^\prime}(\bar{k}_l-\bar{k}_{l^\prime})^2(k_m-k_{m^\prime})^2.
\end{aligned}
\end{equation}
where $m,m^\prime\in\{r_2,\cdots,r_\sigma\};l,l^\prime\in\{j_1,j_2,\cdots,j_\sigma\}$ and $\hat{\sigma}=\{1,2,\cdots,N\}\setminus\{r_2,\cdots,r_\sigma\}$
denotes a subset of the set $\{1,2,\cdots,N\}$.
It is remarked that the $N$-soliton solution of the AB equations can be obtained from (\ref{d5}),(\ref{d10}) and
(\ref{d25})-(\ref{d30}).

\setcounter{equation}{0}
\section{\label{sec:five}Asymptotic behaviors of the $N$-soliton solution}
In this section, we discuss the asymptotic behaviors of the given $N$-solion solution. To this end, we assume that
$$1<|k_1|<|k_2|<\cdots<|k_N|,\quad {\rm Im}k_j>0.$$
It is noted that
\begin{equation}\label{e1}
g_j=e^{2iz_j}=e^{2\theta_j}e^{-2i\varphi},\quad \theta_j={\rm Im}k_j(\xi-v_j\tau-\xi_j),
\end{equation}
where $v_j=-|k_j|^{-2}$ and $\xi_j$ is a certain real constant.
Now the region of the point $\xi=\xi_j+v_j\tau$ is denoted by $\Sigma_j$. Then, as $\tau\rightarrow-\infty$, these regions
are disjoint and distribute from left to right as
$$\Sigma_N,~\Sigma_{N-1},~\cdots,~\Sigma_1,$$
in view of $v_1<v_2<\cdots<v_N$.
In the region $\Sigma_j$, one may find that
\begin{equation}\label{e2}
\begin{aligned}
\xi-\xi_n-v_n\tau\rightarrow+\infty,\\
|g_n|\rightarrow+\infty,\quad n>j;
\end{aligned}
\end{equation}
and
\begin{equation}\label{e3}
\begin{aligned}
\xi-\xi_m-v_m\tau\rightarrow-\infty,\\
|g_m|\rightarrow0,\quad m<j.
\end{aligned}
\end{equation}
Thus, in the region $\Sigma_j$, as $\tau\rightarrow-\infty$, we find
\begin{equation}\label{e4}
\begin{aligned}
&\det(I+M)\approx K\left(\begin{aligned}
j+1,j+2,\cdots,N\\
j+1,j+2,\cdots,N
\end{aligned}\right)\bar{K}\left(\begin{aligned}
j+1,j+2,\cdots,N\\
j+1,j+2,\cdots,N
\end{aligned}\right)\\
&\qquad+K\left(\begin{aligned}
j,j+1,\cdots,N\\
i,j+1,\cdots,N
\end{aligned}\right)\bar{K}\left(\begin{aligned}
j,j+1,\cdots,N\\
j,j+1,\cdots,N
\end{aligned}\right)\\
&=\left(1+\frac{e^{4\theta_j}}{|k_j-\bar{k}_j|^2}\prod\limits_{l=j+1}^N\frac{|k_j-k_l|^4}{|k_j-\bar{k}_l|^4}\right)
\prod\limits_{j+1\leq l<l^\prime\leq N}\frac{e^{4\theta_l}}{|k_l-\bar{k}_l|^2}\frac{|k_l-k_{l^\prime}|^4}{|k_l-\bar{k}_{l^\prime}|^4},
\end{aligned}
\end{equation}
and
\begin{equation}\label{e5}
\begin{aligned}
\det(I&+M+g^TE)-\det(I+M)\\
&\approx G\left(\begin{aligned}
j,j+1,\cdots,N\\
0,j+1,\cdots,N
\end{aligned}\right)H\left(\begin{aligned}
0,j+1,\cdots,N\\
j,j+1,\cdots,N
\end{aligned}\right)\\
&=e^{2\theta_j}e^{-2i\varphi_j}\prod\limits_{l=j+1}^N\frac{(k_j-k_l)^2}{(k_j-\bar{k}_l)^2}
\prod\limits_{j+1\leq l<l^\prime\leq N}\frac{e^{4\theta_l}}{|k_l-\bar{k}_l|^2}\frac{|k_l-k_{l^\prime}|^4}{|k_l-\bar{k}_{l^\prime}|^4}.
\end{aligned}
\end{equation}
Equations (\ref{e4}) and (\ref{e5}) imply that the solution $A$ in $\Sigma_j$ has the following asymptotic behavior
\begin{equation}\label{e6}
A\approx -i{\rm Im}k_je^{-2i(\varphi_j+\delta_j^{(-)})}{\rm sech}2(\theta_j+\gamma_j^{(-)}),\quad \tau\rightarrow-\infty,
\end{equation}
where $\delta_j^{(-)}$ and $\gamma_j^{(-)}$ are defined by the following representation
\begin{equation}\label{e7}
\frac{1}{\bar{k}_j-k_j}\prod\limits_{l=j+1}^N\frac{(\bar{k}_j-\bar{k}_l)^2}{(\bar{k}_j-k_l)^2}=e^{2(\gamma_j^{(-)}+i\delta_j^{(-)})}.
\end{equation}
Similarly, as $\tau\rightarrow+\infty$, the regions are distributed as $\Sigma_1,\cdots,\Sigma_N$.
In this case, the leading term of $\det(I+M)$ in (\ref{e4}) will involve $\{1,\cdots,j-1; 1,\cdots,j-1,j\}$, instead of
$\{j+1,\cdots,N; j,j+1,\cdots, N\}$. Also, in the leading term of (\ref{e5}) is now $\{1,2,\cdots,j-1,j;0,1,\cdots,j-1\}$.
Thus in the region $\Sigma_j$,  we find
\begin{equation}\label{e8}
A\approx -i{\rm Im}k_je^{-2i(\varphi_j+\delta_j^{(+)})}{\rm sech}2(\theta_j+\gamma_j^{(+)}),\quad \tau\rightarrow+\infty,
\end{equation}
and
\begin{equation}\label{e9}
\frac{1}{\bar{k}_j-k_j}\prod\limits_{l=1}^{j-1}\frac{(\bar{k}_j-\bar{k}_l)^2}{(\bar{k}_j-k_l)^2}=e^{2(\gamma_j^{(+)}+i\delta_j^{(+)})}.
\end{equation}

From (\ref{e6}) to (\ref{e9}), one may find that the $N$-solitons with different velocity split as $\tau\rightarrow-\infty$, after mutual collisions,
split again as $\tau\rightarrow+\infty$. In this solitary wave collisions, the form and velocity of each solitary wave do not change, only the center
and phase change from $\delta_j^{(-)},\gamma_j^{(-)}$ to $\delta_j^{(+)},\gamma_j^{(+)}$ for the $j$-th soliton.

Similar considerations apply to $\psi_{21}(0)$ and $\psi_{22}(0)$, we find
\begin{equation}\label{e10}
\psi_{21}(0)\approx\left\{\begin{aligned}
&e^{-i2(\varphi_j+\varrho_j^{(-)})}\sin\varpi_j{\rm sech}2(\theta_j+\gamma_j^{(-)}),&\tau\rightarrow-\infty,\\
&e^{-i2(\varphi_j+\varrho_j^{(+)})}\sin\varpi_j{\rm sech}2(\theta_j+\gamma_j^{(+)}),&\tau\rightarrow+\infty,\\
\end{aligned}\right.
\end{equation}
where $\varrho_j^{(\pm)}=\delta_j^{(\pm)}+\rho_j^{(\pm)}+\frac{\varpi_j}{2}, \varpi_j=\arg k_j,
\rho_j^{(-)}=\sum\limits_{l=j+1}^N\varpi_l, \rho_j^{(+)}=\sum\limits_{l=1}^{j-1}\varpi_l$, 
and
\begin{equation}\label{e11}
\psi_{22}(0)\approx\left\{\begin{aligned}
&1-e^{-i2(\varphi_j+\tilde\varrho_j^{(-)})}e^{2(\theta_j+\mu_j^{(-)})}{\rm sech}2(\theta_j+\gamma_j^{(-)}),&\tau\rightarrow-\infty,\\
&1-e^{-i2(\varphi_j+\tilde\varrho_j^{(+)})}e^{2(\theta_j+\mu_j^{(+)})}{\rm sech}2(\theta_j+\gamma_j^{(+)}),&\tau\rightarrow+\infty,\\
\end{aligned}\right.
\end{equation}
with $\varrho_j^{(\pm)}=\delta_j^{(\pm)}+\nu_j^{(\pm)}$ and $\mu_j^{(\pm)},\nu_j^{(\pm)}$ defined by
$$\begin{aligned}
\frac{1}{2\bar{k}_j}\prod\limits_{l=j+1}^N\frac{\bar{k}_j-\bar{k}_l}{\bar{k}_j-k_l}
\prod\limits_{l=j+1}^N\frac{k_j(k_j-\bar{k}_l)(\bar{k}_j-\bar{k}_l)}{\bar{k}_j(\bar{k}_j-k_l)(k_j-k_l)}=e^{2(\mu_j^{(-)}+i\nu_j^{(-)})},\\
\frac{1}{2\bar{k}_j}\prod\limits_{l=1}^{j-1}\frac{\bar{k}_j-\bar{k}_l}{\bar{k}_j-k_l}
\prod\limits_{l=1}^{j-1}\frac{k_j(k_j-\bar{k}_l)(\bar{k}_j-\bar{k}_l)}{\bar{k}_j(\bar{k}_j-k_l)(k_j-k_l)}=e^{2(\mu_j^{(+)}+i\nu_j^{(+)})}.
\end{aligned}$$
Hence, the asymptotic behavior of the solution $B$ can be characterized by (\ref{d5}) and (\ref{e10}), (\ref{e11}), and
the solitary wave collisions can be discussed similarly.

\section{Conclusions and remarks}
It is remarked that the $\bar\partial$-approach is starting from the dispersion relations of the AB system, which are introduced in
linear equations of the spectral transform matrix $R$ of the $\bar\partial$-problem. By virtue of the $\bar\partial$-dressing method,
we obtain two linear spectral problems, which reduce to the Lax pair of the AB system by using of the associated symmetry conditions.
We note that these symmetries about potential $Q$ and eigenfunction $\psi$ play a crucial role in the determination of the form of spectral transform matrix $R$.
The solutions in closed form, including soliton solutions, are obtained by virtue of the algebraic approach.
%

From section 5, we find that the envelope solitary wave is $v_R\equiv-1/|k_j|^2$, and the velocity of the carrier wave is $v_I\equiv1/|k_j|^2$.
Furthermore, the peculiarity of present solitons is that the center and the phase difference of solitons are dependent on the discrete spectrum,
which is determined by the symmetry conditions of AB system. In addition, the present solitons are stable for $|k_j|>1$ by the results in \cite{G-J-M,G-M,T-B},
for the reason that the velocity of the envelope solitary wave $v_R$ admits $-1<v_R<0$.
\vskip5mm
\noindent{\bf Acknowledgments}\\

Project 11001250 and 10871182 were supported by the National Natural Science
Foundation of China. 


\begin{thebibliography}{99}
\small
\bibitem{PJ70}
Pedlosky J 1970 Finite-amplitude baroclinic waves \textit{J. Atmos. Sci.} \textbf{27} 15-30
\bibitem{PJ}
Pedlosky J 1972 Finite amplitude baroclinic wave packets \textit{J. Atmos. Sci.} \textbf{29} 680-6
\bibitem{MI}
Moroz I M 1981 Slowly modulated baroclinic waves in a three-layer model \textit{J. Atmos. Sci.} \textbf{38} 600-8
\bibitem{M-B}
Moroz I M and Brindley J 1981 Evolution of baroclinic wave packets in a flow with continuous shear and stratification \textit{Proc. Roy. Soc. London A} \textbf{377} 397-404
\bibitem{D-E-G-M}
Dodd R K, Eilbck J C, Gibbon J D and Morris H C 1982 \textit{Solitons and Nonlinear Wave Equations} (New York Academic)
\bibitem{G-J-M}
Gibbon J D, James I N and Moroz I 1979 An example of soliton behavior in a rotating baroclinic fluid \textit{Proc. Roy. Soc. London A} \textbf{367} 219-37
\bibitem{G-M}
Gibbon J D and McGuiness M J 1981 Amplitude equations at the critical points of unstable dispersive physical systems
\textit{Proc. Roy. Soc. A} \textbf{337} 185-219
\bibitem{K-P}
Kamchatnov A M and Pavlov M V 1995 Periodic solutions and Whitham equations for the AB system
\textit{J. Phys. A: Math, Gen.} \textbf{28} 3279-88
\bibitem{T-B}
Tan B and Boyd J P, 2002 Envelope solitary waves and periodic waves in the AB equationss \textit{Stud. Appl. Math.} \textbf{109} 67-87
\bibitem{G-T}
Guo R and Tian B 2012 Integrability aspects and soliton solutions for an inhomogeneous nonlinear system with symbolic computation
\textit{Commun. Nonlinear Sci. Numer. Simulat} \textbf{17} 3189-203
\bibitem{Z-M}
Zakharov V E and Manakov S V 1985 The construction of multidimensional nonlinear integrable systems and their solutions
\textit{Func. Anal. Appl.} \textbf{19} 89-101
\bibitem{B-M}
Bogdanov L V and Manakov S V 1988 The nonlocal $\bar\partial$-problem and (2+1)-dimensional soliton equations,
\textit{J. Phys. A: Math. Gen.} \textbf{21} L537-44
\bibitem{BC}
Beals R and Coifman R R 1989 Linear spectral problems, non-linear equations and the $\bar\partial$-method,
\textit{Inverse Problems} \textbf{5} 87-130
\bibitem{Z}
Zakharov V E, 1990 \textit{On the Dressing Method}, in \textit{Inverse Problems in Action} (ed. Sabatier P S, Springer-Verlag, Berlin) 602-23
\bibitem{SPM}
Santini P M 2003 \textit{Transformations and reductions of integrable nonlinear equations and the $\bar\partial$-problem}
Geometry And Integrability, Ed.Lionel Mason, Yavuz Nutku, (Cambridge University Press)
\bibitem{KBG}
Konopelchenko B G 1993  \textit{Solitons in Multidimensions} (World Scientific, Singapore)
\bibitem{DL}
Doktorov E V and Lebel S B 2007 \textit{A Dressing Method in Mathematical Physics} Springer
\bibitem{Z-G}
Zhu J Y and Geng X G 2012 A hierarchy of coupled evolution equations with self-consistent sources and the dressing method, \textit{J. Phys. A: Math. Theor.} \textbf{46} 035204
\bibitem{H}
Huang N N, 1996 \textit{Theory of Solitions and Method of Perturbations}, (Shanghai Scientific and Technological Education Publishing House,
SHANGHAI) (In Chinese).
\end{thebibliography}

\end{document}